\documentclass[journal]{IEEEtran}
\usepackage{bm}
\usepackage{cite}
\usepackage[pdftex]{graphicx}
\usepackage{multirow}
\usepackage{caption}
\usepackage{subcaption}
\usepackage{multicol}
\usepackage{comment}
\usepackage{mhchem}
\usepackage{color}
\newsavebox{\bigimage}
\usepackage{soul}

\begin{document}

\title{Climate Change Sensing through\\Terahertz Communications:\\A Disruptive Application of 6G Networks}


\author{Lasantha~Thakshila~Wedage, Bernard~Butler, \thanks{\it{Lasantha Thakshila Wedage and Bernard Butler are with the Walton Institute, Waterford Institute of Technology.}} Sasitharan~Balasubramaniam, \thanks{\it{Sasitharan Balasubramaniam is with the University of Nebraska-Lincoln.}} Yevgeni~Koucheryavy, \thanks{\it{Yevgeni Koucheryavy is with the Tampere University of Technology.}} and Josep~M.~Jornet \thanks{\it{Josep Miquel Jornet is with the Northeastern University.}}}

\maketitle

\begin{abstract}
Climate change resulting from the misuse and over-exploitation of natural resources has affected and continues to impact the planet's ecosystem. This pressing issue is leading to the development of novel technologies to sense and measure damaging gas emissions. In parallel, the accelerating evolution of wireless communication networks is resulting in wider deployment of mobile telecommunication infrastructure. With 5G technologies already being commercially deployed, the research community is starting research into new technologies for 6G. One of the visions for 6G is the use of the terahertz (THz) spectrum. In this paper, we propose and explore the use of THz spectrum simultaneously for ultrabroadband communication and atmospheric sensing by leveraging the absorption of THz signals. Through the use of machine learning, we present preliminary results on how we can analyze signal path loss and power spectral density to infer the concentration of different climate-impacting gases. Our vision is to demonstrate how 6G infrastructure can provide sensor data for climate change sensing, in addition to its primary purpose of wireless communication. 
\end{abstract}


%
\IEEEpeerreviewmaketitle

\section*{Introduction}
{\label{1}}

\IEEEPARstart{C}{limate} change is one of the most pressing challenges for humanity and the sustainability of the planet in the twenty-first century. Such challenges include a rise in global temperature that is leading to warmer oceans and shrinking ice sheets contributing to rising sea-levels and ocean acidification. The impact of all these changes on the planet is being witnessed today through more frequent extreme weather events. Researchers believe that the current global climate trend will worsen significantly in the coming decades with increasing greenhouse gas concentrations resulting from human activities relating to expanding industries, new technologies and agricultural activities. All these activities result in an increase of greenhouse effect gases such as carbon dioxide (\ce{CO2}), methane (\ce{CH4}), and nitrous oxide (\ce{N2O}), among others. These gases allow sunlight to pass through the Earth's atmosphere but trap the resulting heat near the surface, which contributes to global warming. Therefore, sensing greenhouse gases as well as other harmful gases (e.g., ozone) can allow the current generation to act and plan for the future by developing new strategies to reduce their emission. Rather than using conventional sensor networks to detect these gases, the question is whether other novel sensing techniques can be developed, without requiring a) massive deployment effort and costs, b) ongoing maintenance and c) material resources (in the form of specialised sensing infrastructure).
  
While the telecommunications industry is rolling out 5G globally, the research community has begun researching new disruptive technologies for future 6G. One of the key technologies for 6G is progression into the upper millimeter-wave (100-300~GHz) and the terahertz (0.3-10~THz) spectrum. The \emph{larger bandwidth available at THz frequencies} (up to hundreds of contiguous GHz) has the potential to provide high data rates that can go up to a terabit-per-second (Tbps) or more. The \emph{shorter wavelength of the THz spectrum} (less than a millimeter) enables both the creation of miniature antennas for nanoscale machine communication in nanonetworks, as well as, through the integration of many such antennas into high-density antenna arrays, the design of highly directional THz links with low probability of detection and interception~\cite{s2:14}. Beyond communications, the combination of very short wavelengths with the \emph{higher photon energies of THz radiation} (though still lower than that of optical signals) improves the resolution and accuracy of traditional radar systems and enables new sensing techniques, including spectroscopy-based classification of media~\cite{s2:13}. Indeed, several frequencies in the THz band are known to be strongly impacted by molecular absorption, and, thus, traditionally, communication systems have avoided those frequencies. However, by changing our perspective, molecular absorption at THz frequencies is also the enabler of atmospheric sensing technologies~\cite{Absorption}. For example, there are multiple satellites orbiting the Earth with THz sensors used for atmospheric studies.

\begin{figure*}[!ht]
    \centering
    \includegraphics[width=1.0\textwidth,height=10cm]{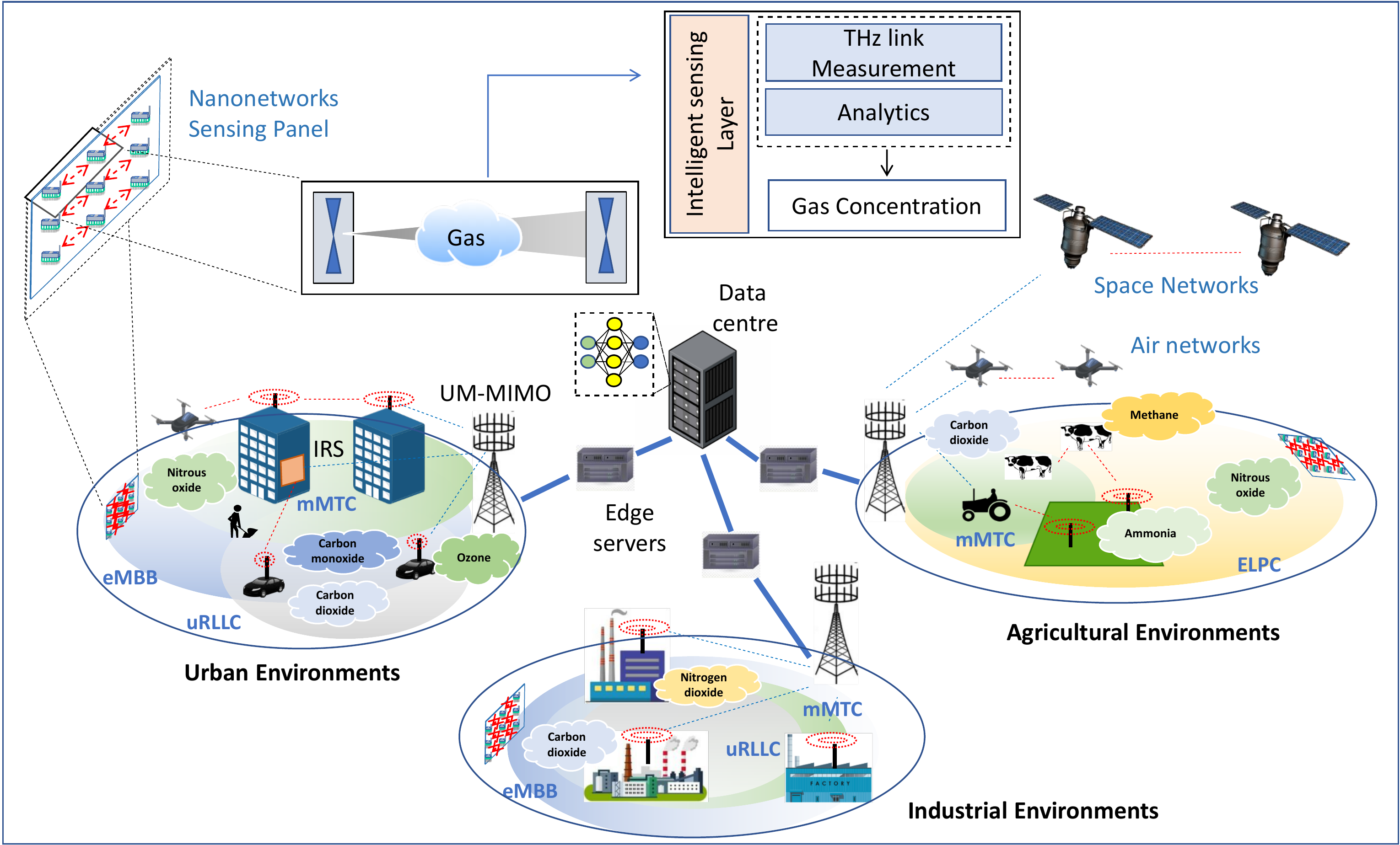}
    \caption{6G Network architecture for communication and sensing}
    \label{Fig1}
\end{figure*}

In our vision, 6G systems have the opportunity to integrate communications and sensing in a totally different way than at lower frequencies, where joint communications and sensing usually means joint communications and radar. More importantly, innovative 6G infrastructure can simultaneously satisfy the connectivity needs of an hyper-connected society while simultaneously collecting an unprecedented amount of data to monitor and, eventually, stop and revert climate change. In Fig.~\ref{Fig1}, we illustrate our envisioned 6G architecture for joint communications and sensing of atmospheric gases.  

Artificial Intelligence (AI)-integrated communication technologies of enhanced Mobile Broadband (eMBB), ultra-Reliable Low Latency Communication (uRLLC), massive Machine Type Communication (mMTC), and Extremely Low Power Communication (ELPC) will utilize novel infrastructures, such as ultra-massive MIMO transceivers, intelligent reflecting surfaces (IRS) based on novel plasmonic reflect-arrays or metasurfaces, and even non-invasive pervasive deployments of nanonetworks. The increased antenna arrays in UM-MIMO coupled with IRS will result in a spatial blanket of THz signals covering the environment. 

Beyond highly-anticipated applications (e.g., automated driving, holography, tactile and haptic internet) and new forms of connectivity (e.g., UAVs), such infrastructure will also enable for the first time distributed atmospheric sensing for climate change monitoring, pollution, and air quality control. In all these, machine learning (ML) and AI will play a key role to analyze the massive amount of collected data and unveil trends and realities.
 
In this paper, we explore for the first time this vision and discuss how THz signal analysis can be used to infer and determine changes in gas concentration that impacts on the climate, which will open new opportunities to gather sensor data of atmospheric gases from telecommunication infrastructure. The remainder of this paper is organized as follows. In the next section, we overview the current sub-terahertz and terahertz technologies used for gas sensing. Then, we present our proposed 6G infrastructures that can be used for sensing gases for various environments. After that, we present preliminary results relating to the use of ML to extract sensing information from the analysis of path loss and received signal power spectral density. Finally, in the last two sections, we identify the challenges that need to be addressed to enable this transformative paradigm and conclude the paper.

\section*{Sub-Terahertz and Terahertz Gas Sensing Technologies}
{\label{2}}
Many gases emitted from agricultural, manufacturing, and industrial processes, as well as urban environments more generally, are harmful pollutants and contribute to the greenhouse effect. Interestingly, in most of these cases, the gases can be detected using THz spectroscopy. We summarize the next state of the art related to THz sensing technologies that can be used to sense toxic, pollutant, and greenhouse gases. 

\subsection*{Sensing for Agricultural Environments}
{\label{2.1}}
Ammonia (\ce{NH3}) is a gas that is found extensively in farming environments, being released by the breakdown of artificial fertilizers and animal manure. Excessive exposure to \ce{NH3} can negatively impact environmental biodiversity. In \cite{t1}, THz frequencies have been used for \ce{NH3} gas and water vapor (\ce{H2O}) sensing using THz Time-Domain Spectroscopy (TDS) transmission measurement geometry. Plants and vegetables are also known to emit Volatile Organic Compounds (VOCs) from leaves, where they are found to enhance crop productivity and ensure food security by inhibiting the germination and growth of pathogens. THz wave spectrometry has been used for VOC gas sensing  \cite{t1:4}. The gases from VOCs such as acetonitrile, ethanol, and methanol (\ce{CH3OH}) can have adverse effects on the human body, and in certain cases can transform into more harmful molecules through chemical reactions within the human body. As an example, acetonitrile can transform to cyanide within the body. 

\subsection*{Sensing for Industrial and Urban Environments}
{\label{2.2}}
Sulfur dioxide (\ce{SO2}), nitrogen dioxide (\ce{NO2}), and carbon monoxide (\ce{CO}) are known as some of the most prevalent pollutant gases found in the atmosphere. These gases mostly enter the atmosphere when fossil fuels are burnt. \ce{SO2} contributes to pollution through acid rain when reacting with rain. The gas can be detected utilizing micro-core photonic crystal fiber-based gas sensors \cite{SO2}. In a similar way, \ce{NO2} also leads to acid rains through the production of nitric acid. Continuous-wave electronic THz spectrometers can sense \ce{NO2} in the frequency range of 220-330~GHz \cite{NO2}. \ce{CO} is also harmful since it readily displaces oxygen in the bloodstream and can lead to asphyxiation, and the gas can be detected using THz Gas-phase spectroscopy (THz-GPS) in the frequency range of 0.3-1.1~THz \cite{CO}. Plants can naturally produce hydrogen cyanide (\ce{HCN}), which are usually degraded within living organisms to reduced toxicity levels. However, they also result in pollutants based mainly on their use in industries, and an example is the mining industry. Besides \ce{HCN} found in wastewater, it can also be found in gases, and this can be detected using photonic crystal cavity detection techniques at frequencies 1.1-1.3~THz  \cite{t1:3}.

\ce{CO2} emitted by industrial processes and burning of fossil fuels is the dominant but not the only greenhouse gas that is accountable for global climate change. Other gases that are creating greenhouse effects include methane \ce{CH4} , nitrous oxide \ce{N2O}, ozone (\ce{O3}), and Fluorinated gases such as tetrafluoromethane (\ce{CF4}). In \cite{t1:6}, THz spectroscopy is used to detect these gases in an atmospheric simulation chamber using frequency ranges 2-2.7~THz and 0.575–0.625~THz for \ce{CH4}, \ce{CF4}, \ce{N2O} and \ce{O3} respectively. 

\section*{6G for Climate Change Action}
\label{3}
Building on the demonstrated possibility of utilizing THz signals to sense critical gases impacting climate change, in this section, we present innovative 6G THz network infrastructures that can bring the vision of joint communications and gas sensing to reality.

\subsection*{Terahertz and Sub-terahertz Absorption Properties}
\label{3.1}
In addition to the high spreading losses resulting from the very small wavelength of THz signals, which requires the utilization of high gain directional antennas with narrow beams, THz signals are also affected by molecular absorption and, to a lower extent, scattering by dust particles, fog, snowflakes, or rain droplets. The main absorber of THz radiation is water vapor, \ce{H2O}, which has resonances across many THz frequencies leading to extremely high absorption \cite{s2:13}. However, as highlighted in the previous section, THz radiation is absorbed by many gases, including \ce{SO2}, \ce{CO2}, \ce{NH3} and \ce{CH4}. Each gas has its own absorption profile, opening new opportunities for using THz signals for sensing. 
 
\begin{figure}[t]
    \centering 
    \includegraphics[width=\linewidth]{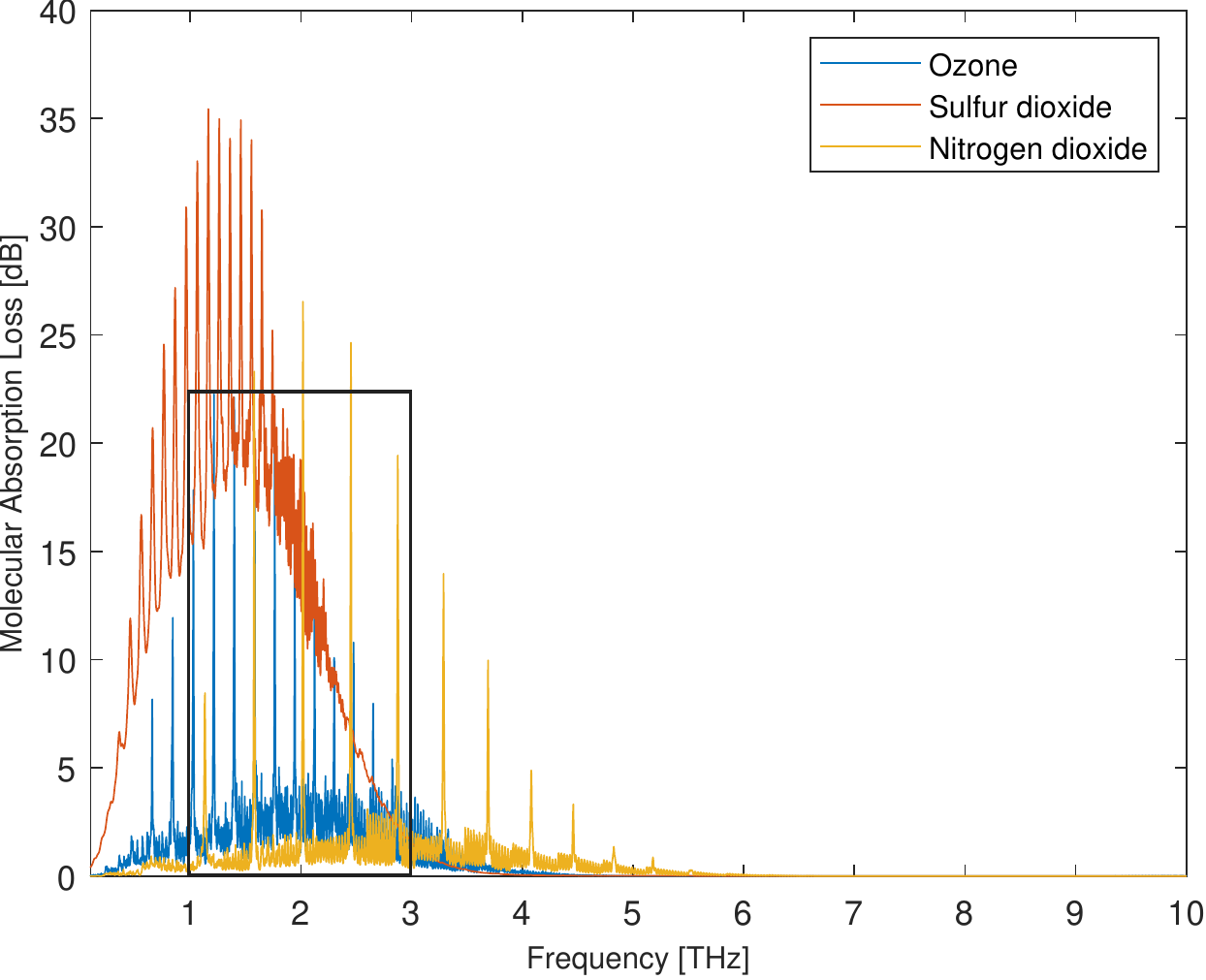}
    \caption{Simulated molecular absorption losses of THz signals for ozone, sulfur dioxide and nitrogen dioxide using HITRAN data}
    \label{Fig2}
\end{figure}

Our sensing concept is based on the molecular absorption profiles, by frequency, that are unique to each gas. Fig.~\ref{Fig2} illustrates simulated molecular absorption losses of THz signals using data from the high-resolution transmission (HITRAN) molecular spectroscopic database \cite{Hitran} under standard temperature (296 K), and pressure (1 atm) for \ce{O3}, \ce{SO2}, and \ce{NO2} when they are mixed with other gases based on their atmospheric concentrations \cite{nasa}. In our measurement model, each standardised absorption profile uniquely identifies a gas, and absorption levels increase with the concentration of that gas. Figure~\ref{Fig2} shows that \ce{SO2} has the highest absorption and so offers greater measurement sensitivity than \ce{NO2} or \ce{O3} over that range of frequencies. 
 
\subsection*{Agricultural Environments}
\label{3.2}
Future farming environments are expected to have multiple sensing devices under the guise of Internet of Everything, communicating to 6G  through mMTC as well as ELPC for Internet of Bio-Nano Things and Internet of Nano Things. The connectivity of these devices can be established through ultra-cells \cite{Ultracell}, which have been proposed for transmitting short-range THz signals. While connectivity from ultra-cell to macrocell will be a problem in rural areas such as farms, the ultra-cells can provide connectivity to local devices and perform edge-based computing, and that can send data to the macrocell (e.g., via drones). In order to redirect beams within farming sheds, which are known to have numerous obstructions due to equipment and facilities, IRS can be used. These ultra-cells will mainly be placed within milking sheds for confined areas to do sensing such as \ce{CH4} emission as cows are being fed, as well as sensing \ce{NH3} and \ce{SO2} from the slurry. Ruminants such as cattle are also known to contribute to greenhouse gases in the form of \ce{CH4} from digesting their food. A single cow emits approximately 200 pounds of \ce{CH4} gas per year. Farm livestock is also known to produce other greenhouse gases such as \ce{CO2} and \ce{N2O}. IRS with ultra-cell-based networking can be utilized to transmit signals in 0.5-1.0~THz frequency to detect a target gas over a distance of more than 1~m, and nanonetwork devices on the IRS itself can be used to sense local gases such as \ce{CH4} and \ce{SO2} over a distance much less than 1~m. Besides deployment within animal sheds, IRS-enabled communication to a mobile vehicle, such as a tractor or drone, can also facilitate gas sensing outdoors.

\begin{table*}[t]

    \centering
    
    \begin{tabular}{|c|p{0.09\linewidth}|p{0.1\linewidth}|p{0.08\linewidth}|p{0.08\linewidth}|p{0.13\linewidth}|p{0.15\linewidth}|c|}
    \hline
    
    & \multicolumn{4}{c|}{Technique using path loss} &\multicolumn{3}{c|}{Techniques developed using spectroscopy} \\
    \hline
 Gas &	Atmospheric concentration (ppm) &	Considered frequency range &	Gaussian noise level &	Possibility of detection &	Frequency range & 	Detection techniques &	Reference\\
 \hline
\ce{H2O}&	10000&	6–8~THz&	1~\%&	Yes&	0.1–2.25~THz&	THz-TDS & \cite{t1}\\
 \hline
\ce{O2}&	209460&	0.5–2.5~THz&	0.01~\%&	Yes& & & \\		
 \hline
\ce{SO2}&	1&	0.5–2.5~THz&	0.01~\%&	Yes	&272.73-333.33~THz &  PCF based gas sensor &\cite{SO2}\\
 \hline
\ce{NH3}&	0.01&	3–5.5~THz&	0.01~\%&	Yes&	0.1–2.25~THz&	THz-TDS & \cite{t1}\\
 \hline
\ce{O3}&	0.07&	1-3~THz&	0.001~\%&	Yes&	0.575–0.625~THz&	THz-TDS & \cite{t1:6}\\
 \hline
\ce{NO2}&	0.02&	1–3~THz&	0.001~\%&	Yes& 0.22-0.33~THz & Continuous-wave electronic THz spectrometer & \cite{NO2}\\		
 \hline
\ce{HCN}&	0.01&	1–3~THz&	0.001~\%&	Yes&	1.1–1.3~THz&	Photonic crystal cavity & \cite{t1:3}\\
 \hline
\ce{CO}&	0.01&	0.5–3~THz&	0.0001~\% &	Yes& 0.3-1.1~THz& THz-GPS& \cite{CO}\\		
 \hline
\ce{CH4}&	1.8&	3-4.5~THz&	0.00001~\%&	Yes&	2–2.7~THz&	THz-TDS & \cite{t1:6}\\
 \hline
\ce{N2}&	780840&	3–5~THz&	Reduced until 0.000001~\%&	No& & & \\		
 \hline
\ce{CO2}&	410&	8–10~THz&	Reduced until 0.000001~\%&	No& & & \\		
 \hline
\ce{N2O}&	0.5&	0.1–1.5~THz&	Reduced until 0.000001~\%&	No&	0.575–0.625~THz&	THz-TDS & \cite{t1:6}\\
 \hline
\ce{CH3OH}&	0.01&	0.1–1~THz&	Reduced until 0.000001~\%&	No&	0.22–0.33~THz&	THz wave  electronics&\cite{t1:4} \\

 \hline
    \end{tabular}
    \caption{ Impact on Gaussian noise level on path loss data analysis for gas concentration measurements.}
    \label{tab:1}
\end{table*}

\subsection*{Industry and Urban Environments}
\label{3.3}
Industrial and urban environments produce greenhouse gases such as \ce{CO2} and \ce{N2O}, mainly as a result of human activities. \ce{CO2} is a major contributor to the global warming crisis. Major industrial sectors producing  \ce{CO2} include power generation (54~percent), cement production (15~percent), gas processing (12~percent), iron refining (6~percent), petroleum refining (5~percent), and chemical plants such as ethanol and ammonia (3~percent) producers. Additionally, large amounts of \ce{CO2} are emitted from residential areas in urban environments as a result of energy consumption. Outdoor infrastructures are most appropriate for sensing these gases. Ultra-massive MIMOs on macrocells, communicating to picocell and femtocells at 0.1-5~THz, can provide opportunities for sensing in industrial and urban environments. Also, using UM-MIMO base stations at 0.3 THz and 1 THz frequency, multi-Tbps links are achievable for communication \cite{UMMIMO}. Moreover, the deployment of femtocell and picocell base stations under the footprint of macrocell base stations reduces the distance between the sensing devices and help to maintain a high signal to interference and noise ratio (SINR) while sensing. Furthermore, picocell base stations are mounted on high-rise buildings or infrastructures in dense urban areas because of their limited coverage  \cite{Ultracell}. Once again, outdoor IRS can also play a significant role in redirecting beams between the cells, and Vehicle to Infrastructure (V2I) communication using the THz links facilitates gas sensing at ground level in urban environments.

\section*{Detection Techniques}
\label{4}
6G will use AI/ML to analyze and process large data sets for its own network management, as well as supporting its use in applications. We consider how AI/ML can be used to infer gas concentrations from measurements of path loss and power spectral density (PSD).

\subsection*{Path Loss Data Analysis}
\label{4.1}

The path loss analysis is based on measuring the attenuation factor and using that information to measure gas concentration. We focus on the molecular absorption loss per frequency rather than the total path loss which includes spreading loss. The spreading loss is based purely on the distance and specific frequency between the transmitter and receiver and so is not affected by the gas mixture. The detection accuracy is determined by the ratio of Gaussian noise to absorption loss in the received signal. Table \ref{tab:1} summarises this for a variety of gases at a ratio specified by the atmospheric concentration, where we can see that each gas type will have a corresponding maximum tolerable Gaussian noise that will result in accurate detection. We also compare the frequency range we used in our study with the frequency range used in THz-spectroscopy based sensing techniques. Our analysis is established by controlling the noise level and reducing it step-by-step to validate a constrained linear least square technique that was used to solve the regression problem with constraints. The conditions in our study include a) the concentrations of each gas should be less than one million ppm, and b) the sum of the concentrations should equal one million ppm. The results in the table are based on 1000 Monte Carlo simulations to estimate the effects of randomness. Our simulations gradually decreased the Gaussian noise level until $0.000001$~percent, and most of the gases in the mixture were detectable and measurable at $0.00001$~percent. The expected atmospheric gas concentrations (in ppm) from Table \ref{tab:1} are used to generate molecular absorption loss profiles for typical atmospheric gas mixtures. Gaussian noise is added to the generated absorption losses in a controlled way, then we try to estimate each gas in the presence of this noise. Some gases, like \ce{H2O}, can be measured even with 1 percent added noise but there is much less sensitivity for gases like \ce{N2O}.

\begin{figure}[t]
    \centering
    \includegraphics[width=\linewidth]{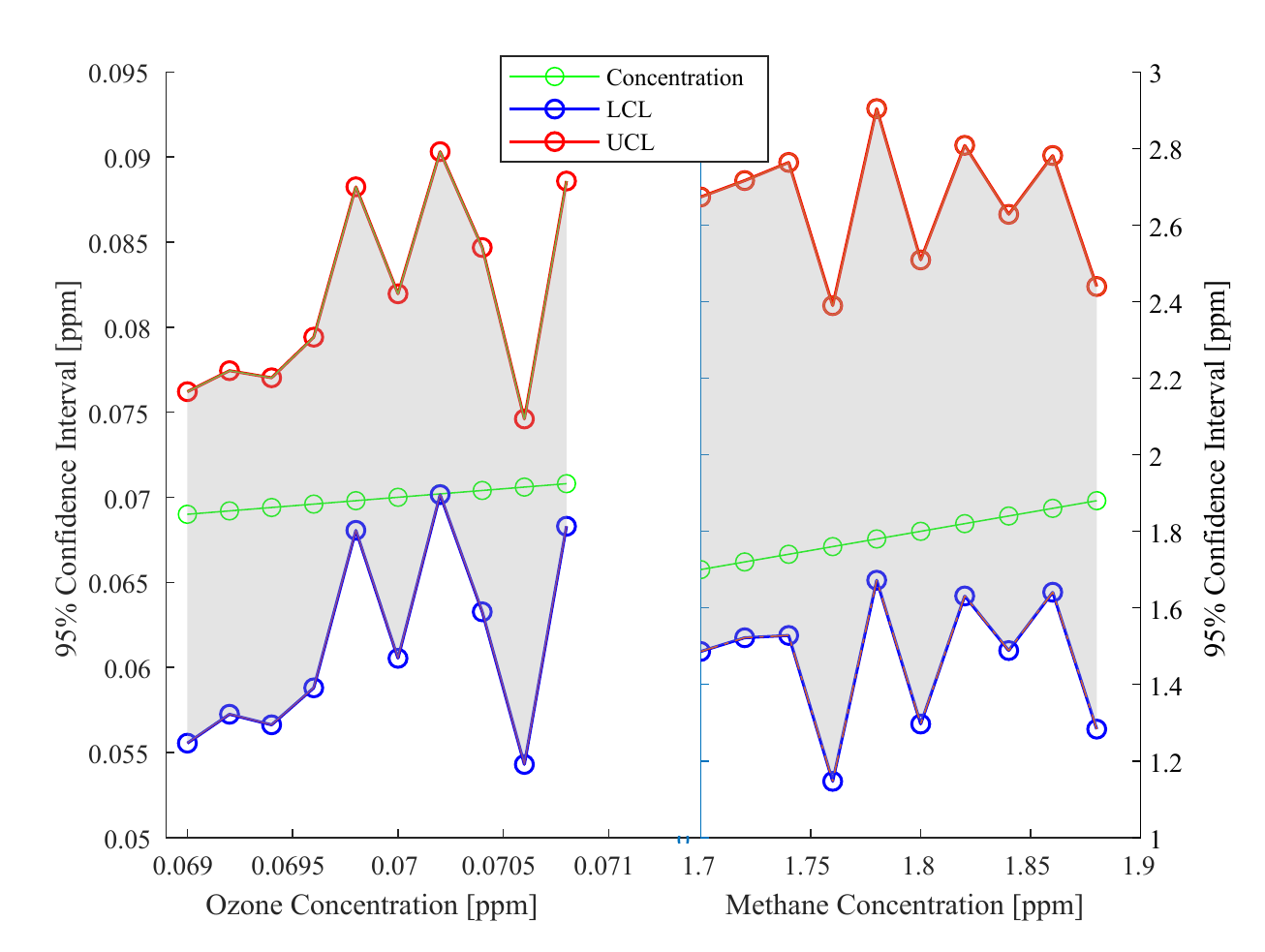}
    \caption{Measurement sensitivity curve for ozone and methane, showing expected gas concentrations and the confidence intervals(LCL: Lower Confidence Level, UCL: Upper Confidence Level) of the predicted gas concentrations.}
    \label{Fig3}
\end{figure}

As an example of using path loss measurement to detect the gas concentration, we applied a multiple linear regression model that considers the total molecular absorption loss is from a linear combination of losses of each individual gas within the mixture.  Figure~\ref{Fig3} shows how we used the linear regression model to predict the concentration of \ce{O3} and \ce{CH4} at 5~cm distance for the frequency range of 1.0-3.0~THz and 3.0-4.5~THz, respectively. The selected frequency range is based on the band with the highest molecular absorption loss for the specific gas type (see the highlighted rectangular area in Fig.~\ref{Fig2} for \ce{O3}). The measurement sensitivity curves for \ce{O3} and \ce{CH4} were generated at the $0.001$~percent and $0.00001$~percent Gaussian noise levels, respectively. The results in Fig.~\ref{Fig3} shows that we can establish 95~percent confidence intervals (CI's) of the predicted gas concentrations that only deviates from the actual concentration by a small percentage (this is bounded by the upper confidence levels (UCL) and lower confidence levels (LCL)). The other gases such as \ce{N2}, \ce{CO2}, \ce{N2O} and \ce{CH3OH} are not measurable using path loss data for any of the considered THz frequencies because the measurement sensitivity is too low.

\begin{figure}[t]
    \centering
    \includegraphics[width=\linewidth]{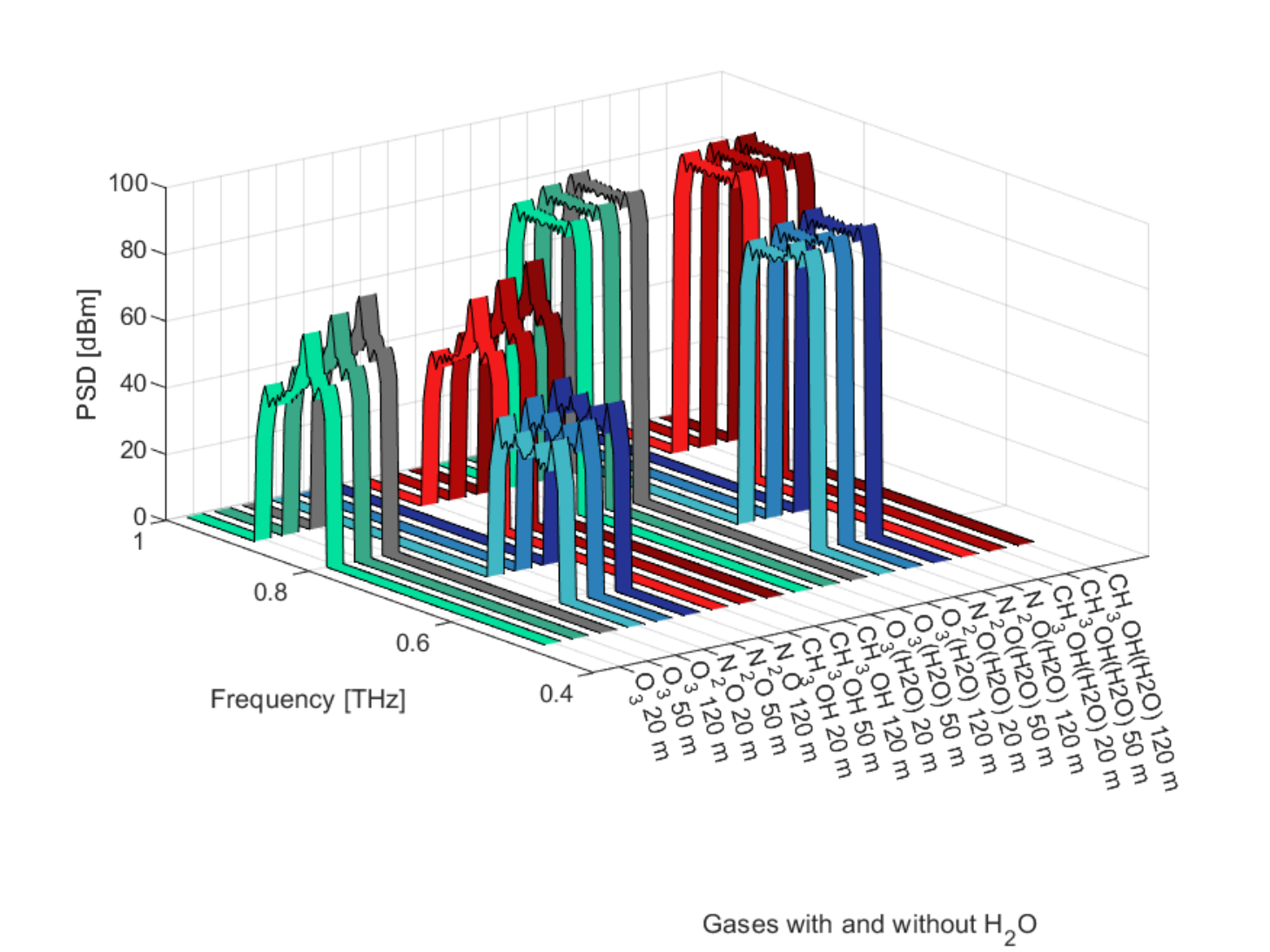}
    \caption{Total power spectral density for ozone, nitrous oxide and methanol considering a mixture with and without water vapor and varying distance between transmitter and receiver.}
    \label{Fig4}
\end{figure}
\subsection*{Power Spectral Density Data Analysis}
\label{4.2}
 We use PSD measurement analysis to sense a targeted gas in a mixture. Figure~\ref{Fig4} presents the power spectral densities of \ce{O3} in 0.59–0.69~THz and \ce{N2O} and \ce{CH3OH} in 0.8–0.9~THz frequency bands by considering a scenario of sending 0.05 nanoseconds long pulsed chirp signals through a gas mixture with and without \ce{H2O}, while also varying the distance between the transmitter and receiver. We analyzed the molecular absorption loss of the targeted gases when mixed with \ce{H2O}, \ce{O2} and \ce{N2} to select the narrow frequency ranges that will result in low absorption loss by \ce{H2O} and high absorption loss for the target gas. Our results show there is a  significant impact from \ce{H2O} on the PSD measurement corresponding to the molecular absorption noise, as well as the attenuation effect of distance. This impact on the overall PSD measurement is summed with the PSD corresponding to the chirp signal in the frequency domain. The shapes in Fig. \ref{Fig4} indicate that it is possible to estimate gas concentrations by applying chirp spread spectrum signals and using supervised-learning techniques.

\begin{figure}[t]
    \centering
    \includegraphics[width=\linewidth]{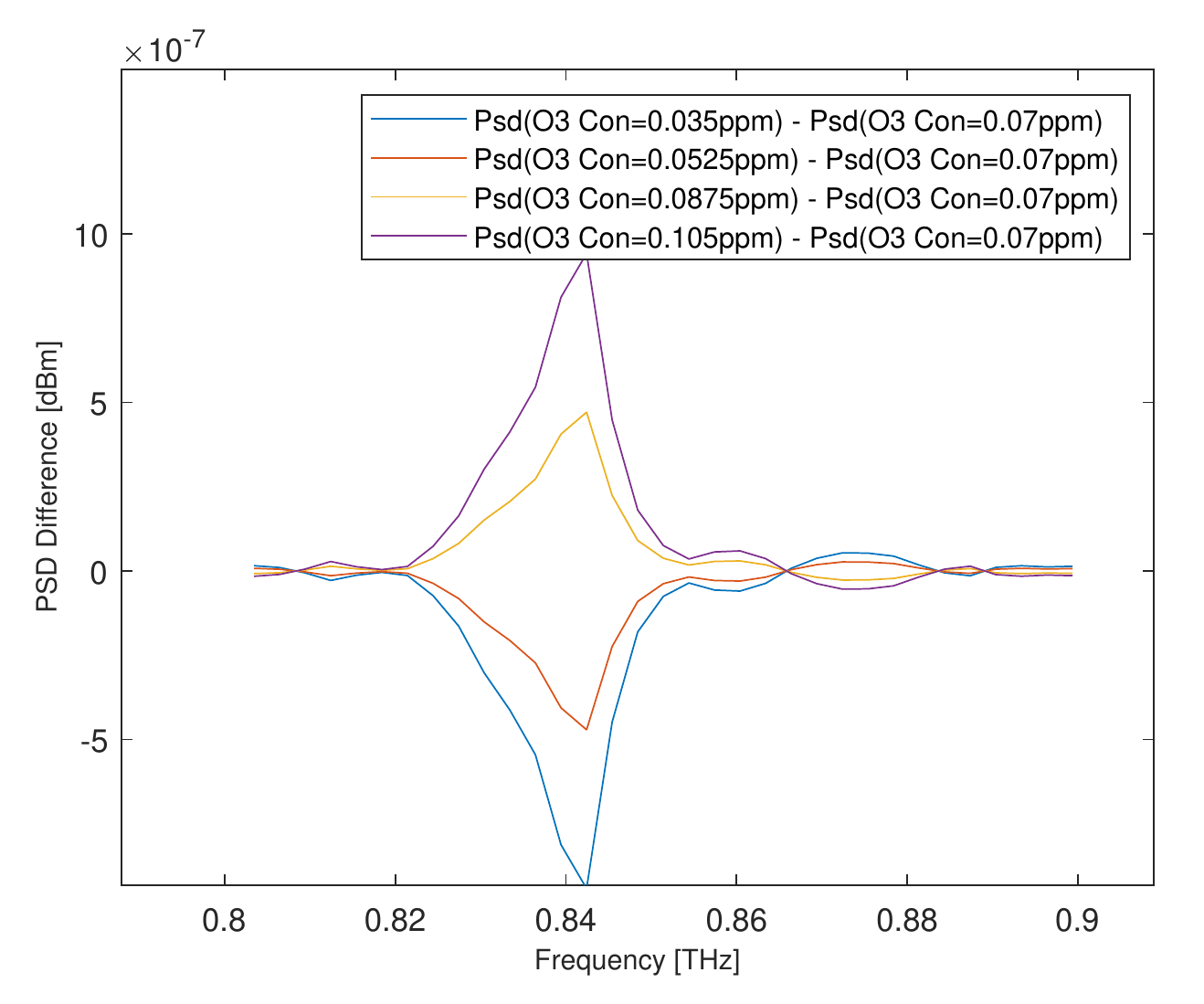}
    \caption{Comparing PSD difference for ozone relative to the standard atmospheric ozone concentration.}
    \label{Fig5}
\end{figure}

Our PSD analysis also considers sensing a target gas when its gas concentration varies in the atmosphere. Typical variation in atmospheric gas concentration is relatively small, so differences in the PSD measurements are very small. To explore this challenge, PSD differences relative to the prevalent atmospheric concentration of the targeted gas were measured for a fixed distance of 100~m between the transmitter and receiver. Figure~\ref{Fig5} presents measurements for \ce{O3}, where we can see a significant difference in the PSD  for all \ce{O3} concentrations at 0.8424 THz frequency. This shows that ML techniques can be applied to locate the changes in PSD quantity to allow us to determine the different gas levels at specific frequencies. In future work, this analysis might be extended to predict a certain gas concentration and localize it using ML Techniques. 

\section*{Challenges}
{\label{5}}
In this section, we list some challenges associated with THz signals that will be used for sensing, and in particular, when deployed onto 6G infrastructure. 

\subsection*{Ultradense Sensing Signals} 
{\label{5.1}}
Given the spatial dispersion of gases within the environment, a significant requirement is the creation of a THz signal blanket that covers an area with sufficient spatial granularity. We could increase infrastructure density, such as IRS and UM-MIMO base stations, to cover specific areas. While drones may be able to carry nanonetwork sensing panels, they might not be able to cover an area for a period long enough to sense the changes in the gas concentration. Therefore, protocols will need to be developed that consider the tradeoffs between maximizing spatial coverage and minimizing energy consumption in order to allow fine-grained spatial sensing. Interference between the beams might occur, but this should not cause serious problems because THz beams are deliberately thin.

\subsection*{Sensing Frequency Switching} 
{\label{5.2}}
A challenge lies in the ability to switch between frequencies on a single device to facilitate communication as well as gas sensing. To minimize the need for switching frequencies between communication and sensing use cases, we plan to investigate sensing using the side-lobes rather than the main lobes of the signals. This in turn might enable communication signals to be used for sensing within the one signal beam. Further investigations are required into metamaterials that can be used to construct different antennas on the unit to enable switching between diverse frequencies. As we have analyzed in our detection techniques, a single frequency signal path loss may not be sufficient, so we may require a chirp spread spectrum that sweeps through multiple frequencies. This provides an opportunity to utilize the large bandwidth in the THz spectrum for sensing a wide frequency range. 

\subsection*{Reconfigurable Beam to Minimize Sensing Deafness} {\label{5.3}}
While (massive) antenna arrays in the THz spectrum can be used to generate pencil-thin beams to overcome the path loss and meet the link budget requirements, different beam widths and beam distributions might be needed to meet the sensing requirements. These include from quasi-omnidirectional short-range beams to single and multiple directional beams. Such flexibility results in hardware challenges that will require on-the-fly reconfiguration of the beam shapes.

\subsection*{Gathering Data for Analysis}
{\label{5.4}}
We propose that path loss as a function of frequency can be used for sensing, but estimating the location of the sensed region remains a challenge. We propose the use of ML to triangulate signals from multiple sources. This will lead to a vast quantity of data for training as well as accurate detection. This data analysis is needed because numerous factors can affect the signals and be confounded with each other, making accurate measurement difficult. The data analysis can also assist in minimizing the energy consumption from each device. This can be achieved by varying the sleep cycles of the sensing duration in line with changes in the measured gas. Our preliminary simulations have shown the use of ML to infer gas concentrations from path loss and PSD. However, further investigations are required to accurately determine the concentrations when \ce{H2O} is present. \ce{H2O} concentration in the atmosphere varies unpredictably due to environmental conditions. Since \ce{H2O} molecular absorption loss is very high compared to the other gases, it is challenging to sense other important gases when the \ce{H2O} percentage is high, e.g., exceeds 1~percent. The atmospheric concentrations of some gases used in the study are expected to be very small, so they are difficult to detect.

\section*{Conclusion}
{\label{6}}
Early visions for 6G systems agree that new infrastructure will be needed in the next generation of wireless systems beyond what is currently being deployed for 5G. Such new infrastructure includes IRS, EM-nanonetworks and increased frequency spectrum in the THz band. In this paper, we have investigated how we can exploit the absorption of THz signals by certain gases as a new sensing technique for 6G communication networks. Through a preliminary machine learning analysis, we have been able to show how path loss and power spectral density can be used to sense various gas types. While many challenges await deployment of our proposed approach, we believe that it can lay the groundwork for research into how newly added functionalities in telecommunication infrastructure can measure data for climate change sensing.

\section*{Acknowledgment}
{\label{7}}

This publication has emanated from research conducted with the financial support of Science Foundation Ireland (SFI) and the Department of Agriculture, Food and Marine on behalf of the Government of Ireland under Grant Number [16/RC/3835] - VistaMilk, and of YL Verkot.

\ifCLASSOPTIONcaptionsoff
  \newpage
\fi

\bibliographystyle{IEEEtran}
\bibliography{References}



\begin{IEEEbiographynophoto}{LASANTHA THAKSHILA WEDAGE} [S'21] (thakshila.wedage@waltoninstitute.ie)
received his B.S. degree in Mathematics from University of Ruhuna, Sri Lanka, in 2016. 
He is currently pursuing a Ph.D. degree with the Department of Computing and Mathematics, Walton Institute, 
Waterford Institute of Technology, Ireland. His current research interests lie in Mathematical modelling and 5G/6G Wireless communication and sensing.
\end{IEEEbiographynophoto}

\begin{IEEEbiographynophoto}{BERNARD BUTLER}[S'10, M'16] (bernard.butler@waltoninstitute.ie) received his PhD degree from Waterford Institute of Technology (WIT), Ireland. He was a Research Scientist in the U.K.’s National Physical Laboratory, focusing on mathematics of measurement and sensing. He is a postdoctoral researcher in the Walton Institute, WIT, where his research interests include the management of distributed computing and sensing systems, applied to future networking, smart cities and agriculture.
\end{IEEEbiographynophoto}

\begin{IEEEbiographynophoto}{SASITHARAN BALASUBRAMANIAM} [SM'14] (sasi@unl.edu) received his PhD degree from the University of Queensland, Australia in 2005. He is current an Associate Professor at the School of Computing, University of Nebraska-Lincoln. His research interests lie in molecular and nano communications, Internet of Bio-Nano Things, as well as 5G/6G.
\end{IEEEbiographynophoto}


\begin{IEEEbiographynophoto}{YEVGENI KOUCHERYAVY}[SM'08] (yevgeni.koucheryavy@yl-verkot.com)
received the Ph.D. degree from the Tampere University of Technology, Finland, in 2004. He is currently a Full Professor with the Unit of Electrical Engineering, Tampere University, Finland. He has authored numerous publications in the field of advanced wired and wireless networking and communications. His current research interests include various aspects in heterogeneous wireless communication networks and systems, the Internet of Things and its standardization, and nanocommunications.
\end{IEEEbiographynophoto}


\begin{IEEEbiographynophoto}{JOSEP MIQUEL JORNET}[M'13,SM'20] (jmjornet@northeastern.edu) received the B.S. and the M.Sc. in Telecommunication Engineering from Universitat Politecnica de Catalunya in 2008, and the Ph.D. degree in Electrical and Computer Engineering (ECE) from Georgia Tech in 2013. Between 2013 and 2019, he was with the Department of Electrical Engineering at University at Buffalo. Since August 2019, he is an Associate Professor in the Department of Electrical and Computer Engineering, the Director of the Ultrabroadband Nanonetworking Laboratory and a Faculty Member of the Institute for the Wireless Internet of Things at Northeastern University. His research interests are in terahertz communications and wireless nano-bio-communication networks.

\end{IEEEbiographynophoto}




\end{document}